\documentclass[aps]{revtex4} 
\usepackage{psfig}
\usepackage{bm}

\begin{document} 
 
\title{Lack of energy equipartition in homogeneous
heated binary granular mixtures}
 
\author{Alain Barrat\footnote{Electronic Address: Alain.Barrat@th.u-psud.fr} 
and Emmanuel Trizac\footnote{Electronic Address: 
Emmanuel.Trizac@th.u-psud.fr}
}

\affiliation{
Laboratoire de Physique Th{\'e}orique
(UMR 8627 du CNRS), B{\^a}timent 210, Universit{\'e} de
Paris-Sud, 91405 Orsay Cedex, France 
}
 
\date{\today}

\begin{abstract}
We consider the problem of determining the granular temperatures
of the components of a homogeneous 
binary heated mixture of inelastic hard spheres,
in the framework of Enskog kinetic theory. Equations are derived
for the temperatures of each species
and their ratio, which is different from unity, as may be expected
since the system is out of equilibrium. We focus on the particular 
heating mechanism where the inelastic energy loss is compensated by 
an injection through a random external force 
(``stochastic thermostat''). The influence of
various parameters and their possible experimental relevance 
is discussed.
\end{abstract} 

\maketitle 

\section{Introduction}

Experimental and theoretical studies of rapid granular flows \cite{Jaeger}
have hitherto
mostly focused on assemblies of identical particles, either
freely cooling when the energy loss due to inter-particle collisions is 
not compensated for, or driven in a non-equilibrium stationary state
by various energy injection mechanisms.
Recently however, interest has grown for the more complicated case of
polydisperse systems 
\cite{Duparcmeur,garzo,Huilin,losert,Wildman,menon,Clelland,MontaneroShear,MontaneroHCS}. 
Theoretical investigations into the homogeneous cooling stage
of a binary mixture \cite{Duparcmeur,garzo} have shown that the
two components have different granular temperatures
(i.e. kinetic energies), even if their cooling rates are
equal. Such a result, confirmed by detailed 
Monte Carlo simulations \cite{MontaneroHCS}
is also obtained  when the system is sheared \cite{MontaneroShear,Clelland},
heated by the contact with an elastic
granular gas maintained at fixed temperature $T$ \cite{biben},
or within the Maxwell model framework \cite{puglisi2}.
Similarly, a tracer particle undergoing inelastic collisions
with an equilibrium fluid at temperature $T$ reaches a granular temperature
lower than $T$ \cite{martin}.

This violation of equipartition in a mixture, although in sharp
contrast with the behaviour of molecular gases at equilibrium, 
is not unexpected: 
the terminology ``granular temperature''
for the kinetic energy of a granular gas has been coined from 
the equivalence of temperature and kinetic energy in an elastic
gas, but does not have any thermodynamical status 
in out-of-equilibrium systems like inelastic granular gases.

In recent experiments, the granular temperatures have been measured
for binary mixtures, both in 3D vibro-fluidized granular beds \cite{Wildman} and 
in 2D strongly vibrated granular gases \cite{menon}. 
Both studies reported a clear violation of equipartition
with a temperature ratio 
quite insensitive to the relative densities of the two species. 

The present article aims at providing a simple theoretical framework 
where the temperature ratio is readily obtained in a 
non-equilibrium steady state (NESS). This allows to investigate 
the influence of many parameters which can be difficult to 
tune experimentally, such as  
the masses, sizes, densities and inelasticities of the beads.
We consider analytically a heated binary mixture in the
framework of the homogeneous non-linear (Enskog-)Boltzmann equation
for smooth inelastic hard spheres. 
Similarly to the case of free cooling described in \cite{garzo}
and restricting to Gaussian velocity distributions,
we derive in section \ref{sec:2} equations for the 
granular temperatures of the mixture components, which
are easily solved numerically. The corresponding temperature 
ratio is in excellent agreement with existing numerical work \cite{biben}.
In section \ref{sec:3}, we consider the NESS sustained by
heating through random kicks (``stochastic thermostat'' approach), 
a mechanism which has focused some attention recently for 
one-component (monodisperse) systems
\cite{Williams,Puglisi,twan,Pre1,Montanero,Cafiero,Moon,Pre2,Garzo}. 
Although finding an energy injection mechanism of experimental relevance
is a difficult issue, we expect the approach proposed here 
to elucidate the basic trends of grain behaviour when varying the
controlling parameters. Moreover, as will be shown below, 
the temperature ratio we obtain provides a reasonable 
zeroth order approximation to compare with the experiments
reported in \cite{Wildman,menon}.

\section{Kinetic theory}
\label{sec:2}

We consider the model of smooth inelastic hard spheres (IHS) undergoing
binary, momentum conserving, inelastic collisions, in the
framework of the homogeneous non-linear Enskog equation. The system
is a mixture of two types of IHS, with masses $m_1$ and $m_2$, diameters
$\sigma_1$ and $\sigma_2$. Three types of collisions may occur so that
the mixture is also characterized by three different restitution coefficients:
$\alpha_{11}$, $\alpha_{22}$, and $\alpha_{12}=\alpha_{21}$.

The velocity distributions in the homogeneous state
$f_1(\bm{v},t)$, $f_2(\bm{v},t)$, obey the following kinetic equations:
\begin{equation}
\partial_t f_i ( \bm{v_1}, t) = \sum_j J_{ij}[ \bm{v_1} | f_i, f_j]
+{\cal F}f_i
\label{eq:Jij}
\end{equation}
where the $J_{ij}$ describe the effect of dissipative inter-particle 
collisions, and ${\cal F } f_i$ represents an external forcing which injects
energy into the system, allowing it to reach a 
non-equilibrium steady state.
The kernels $J_{ij}$ for collisions between
a particle of type $i$ and a particle of type $j$ are given,
in dimension $d$, by
\begin{equation}
J_{ij}[ \bm{v_1} | f_i, f_j] = \chi_{ij} \sigma_{ij}^{d-1}
\int d\bm{v}_2 \int' d\widehat{\bm{\sigma}} 
(\widehat{\bm{\sigma}}\cdot \bm{v}_{12})
\left( \frac{1}{\alpha_{ij}^2}
f_i(\bm{v}_1')f_j(\bm{v}_2') - f_i(\bm{v}_1)f_j(\bm{v}_2) \right)
\ .
\end{equation}
where the $\chi_{ij}$ are the pair distribution functions at contact
(a priori unknown, but becoming close to 1 in the limit of low densities);
$\widehat{\bm{\sigma}}$ is a unit
vector directed from the center of the particle of type $i$ to the
center of particle $j$ [separated at contact by 
$\sigma_{ij}=(\sigma_i+\sigma_j)/2$], and the prime on the integral is
a shortcut for $\int \Theta(\widehat{\bm{\sigma}}\cdot \bm{v}_{12})$.
Moreover, $\bm{v}_{12}= \bm{v}_1 - \bm{v}_2$,
and the pre-collisional velocities $\bm{v}_1'$ and $\bm{v}_2'$
are given in terms of the 
post-collisional velocities $\bm{v}_1$ and $\bm{v}_2$ by:
\begin{eqnarray}
\bm{v}_1'= \bm{v}_1 - \mu_{ji} (1+\alpha_{ij}^{-1})
(\widehat{\bm{\sigma}}\cdot \bm{v}_{12})\widehat{\bm{\sigma}} \\
\bm{v}_2'= \bm{v}_2 + \mu_{ij} (1+\alpha_{ij}^{-1})
(\widehat{\bm{\sigma}}\cdot \bm{v}_{12})\widehat{\bm{\sigma}}
\end{eqnarray}  
where $\mu_{ij}=m_i/(m_i+m_j)$, so that momentum is conserved but
energy dissipated.

The partial granular temperatures are defined from the kinetic energies
by
\begin{equation}
\frac{n_i d}{2} T_i(t) = 
\int d\bm{v} \frac{m_i v^2}{2} f_i(\bm{v},t) \ ,
\end{equation} 
$n_i=\int  d\bm{v}  f_i(\bm{v},t)$ 
being the number density of particles of type $i$ with a total 
temperature of the mixture 
\begin{equation}
T= \frac{1}{n_1+n_2} \sum_i n_i T_i \ .
\end{equation} 

From (\ref{eq:Jij}), the evolution equation for the 
temperatures reads
\begin{equation}
\partial_t T_i = \frac{m_i}{n_i d}
\sum_j \int  d\bm{v} v^2 J_{ij}[ \bm{v} | f_i, f_j] + {\cal FT}_i \ ,
\end{equation}
where ${\cal FT}_i$ describes the effect of the forcing (source) term
${\cal F} f_i$.
It is possible to integrate over $\widehat{\bm{\sigma}}$
(see calculations in the appendix, and also
\cite{garzo}) to obtain, without any further approximation at this
stage:
\begin{eqnarray}
\partial_t T_i &=& {\cal FT}_i 
- \frac{\beta_3 m_i \chi_{ii} \sigma_{ii}^{d-1} (1-\alpha_{ii}^2)}{4 n_i d}
\int d\bm{v}_1 d\bm{v}_2 v_{12}^3 f_i(\bm{v}_1) f_i(\bm{v}_2)
\nonumber \\
&-& \frac{\beta_3 m_i \chi_{ij} \sigma_{ij}^{d-1}}{n_i d}
\int d\bm{v}_1 d\bm{v}_2 f_i(\bm{v}_1) f_j(\bm{v}_2)  
\left[ \mu_{ji}^2 (1- \alpha_{ij}^2)  v_{12}^3
+2 \mu_{ji} (1+\alpha_{ij})  v_{12} (\bm{v}_{12} \cdot \bm{V}_{ij})
\right]
\label{eq:cooling}
\end{eqnarray}
with $\beta_3= \pi^{(d-1)/2}/\Gamma[(d+3)/2]$,
$\bm{V}_{ij}= \mu_{ij} \bm{v}_1 + \mu_{ji} \bm{v}_2$, and 
$\Gamma$ the Euler function.

Since the system reaches a stationary
state where the forcing term balances the dissipation due to collisions
(the forcing and dissipative terms in (\ref{eq:cooling}) generically involve
different powers of the temperatures so that its right-hand-side admits
a ``physical'' root), we can write
$\partial_t T_i = 0$. It is moreover convenient to scale
the velocities with the thermal velocities $v_{0,i}=\sqrt{2T_i/m_i}$,
and introduce the functions $\Phi_i$ such that
\begin{equation}
f_i (v) = \frac{n_i}{v_{0,i}^d} \Phi_i \left( \frac{v}{v_{0,i}} \right) \ .
\end{equation} 

Equation (\ref{eq:cooling}) may then be cast into an equation
for the rescaled velocity distributions $\Phi_i$; no further step can however
be taken without some approximations
on the unknown distributions $\Phi_i$. It is convenient to study the
deviations of $\Phi_i$ from the Gaussian $\Phi_i^0 (c)$ through
an expansion in Sonine polynomials \cite{Landau}. 
In single component heated systems, the deviation from a Gaussian
remains small, especially for experimentally relevant values 
of the restitution coefficient \cite{twan,Montanero,Moon,Pre2}. 
We will here limit our treatment to the Gaussian approximation 
$\Phi_i (c) =\Phi_i^0 (c) = \pi^{d/2} \exp(-c^2)$
(lowest order Sonine expansion). It would of course
be possible to go further in a systematic and controlled way, 
as in \cite{garzo}, but we will
see by comparison of our approximate analytical calculations 
with Monte Carlo simulations
that, at least in the cases we consider,
the Gaussian approximation provides reliable results.

Assuming Gaussian velocity distributions,  
it is now possible to carry out the remaining integrations
in (\ref{eq:cooling}); the calculations
are straightforward and some technical details may be found 
in the appendix of \cite{garzo}. We only give the resulting
equations for the granular temperatures $T_i$ in the NESS:
\begin{eqnarray}
\frac{d \Gamma(d/2)}{m_i \pi^{(d-1)/2}}{\cal FT}_i &=&
\chi_{ii} \sigma_{ii}^{d-1} n_i \frac{2(1-\alpha_{ii}^2)}{m_i^{3/2}} T_i^{3/2}
+ \chi_{ij} \sigma_{ij}^{d-1} n_j \mu_{ji}
\left[ \mu_{ji} (1-\alpha_{ij}^2)
\left( \frac{2T_i}{m_i} +  \frac{2T_j}{m_j} \right) \right.
\nonumber \\
&+& \left. 4 (1+\alpha_{ij})\frac{T_i - T_j}{m_1+m_2} \right]
\left( \frac{2T_i}{m_i} +  \frac{2T_j}{m_j} \right)^{1/2} \ .
\label{eq:T1T2}
\end{eqnarray}
These equations still depend on the particular heating mechanism
through the term ${\cal FT}_i$; once the latter has been specified,
two equations are obtained for $T_1$ and $T_2$; they are easy to implement and
solve numerically varying the various controlling parameters.

Before turning to the heating provided by a stochastic thermostat
[which amounts to $({\cal FT}_i) / m_i = \hbox{constant}$],
we consider three particular limiting cases.

\paragraph{}
In the tracer limit \cite{martin}, i.e. 
$n_1 \to 0$, without any forcing term, $T_2$ is imposed and only the equation
for $T_1$ is considered. As already noted in \cite{garzo}, the result 
for $\gamma=T_1/T_2$ obtained in \cite{martin}
\begin{equation}
\gamma = \frac{1+\alpha_{12}}{2 + \frac{m_2}{m_1}(1-\alpha_{12})}
\end{equation}
is easily recovered, irrespective of dimension. 

\paragraph{}
Another possibility to obtain a NESS for IHS
has been proposed in \cite{biben}: the temperature
$T_2=T$ of one population is imposed, with a corresponding 
Gaussian velocity distribution, and elastic collisions
between particles of type $2$ as well as for $1-2$ collisions:
$\alpha_{22}=\alpha_{12}=1$. Energy is consequently injected into
the inelastic population $1$ (with restitution coefficient 
$\alpha_{11}=\alpha < 1$), without the need for any other forcing term.
In \cite{biben}, high precision
numerical results were obtained for the distribution function $\Phi_1$, and
temperature $T_1$ (the system is three-dimensional, and $\chi_{ij}=1$), by
an iterative numerical resolution of the Boltzmann equation. 
Imposing ${\cal F}T_1=0$ in (\ref{eq:T1T2}), it is straightforward to obtain 
a third order polynomial equation for $\gamma=T_1/T_2$ (this quantity
is necessarily smaller than 1)
\begin{equation}
\frac{\epsilon^2 (1-\alpha^2)^2}{32}
\left( \frac{m_1}{m_2} + \frac{m_2}{m_1} +2 \right)^2 \gamma^3
= (1-\gamma)^2 \left( \gamma+ \frac{m_1}{m_2} \right) \qquad \hbox{where} \qquad
\epsilon = \frac{4n_1}{n_2(1+\sigma_2/\sigma_1)^2}. 
\label{eq:biben}
\end{equation}

\begin{figure}[htb]
\centerline{
\psfig{figure=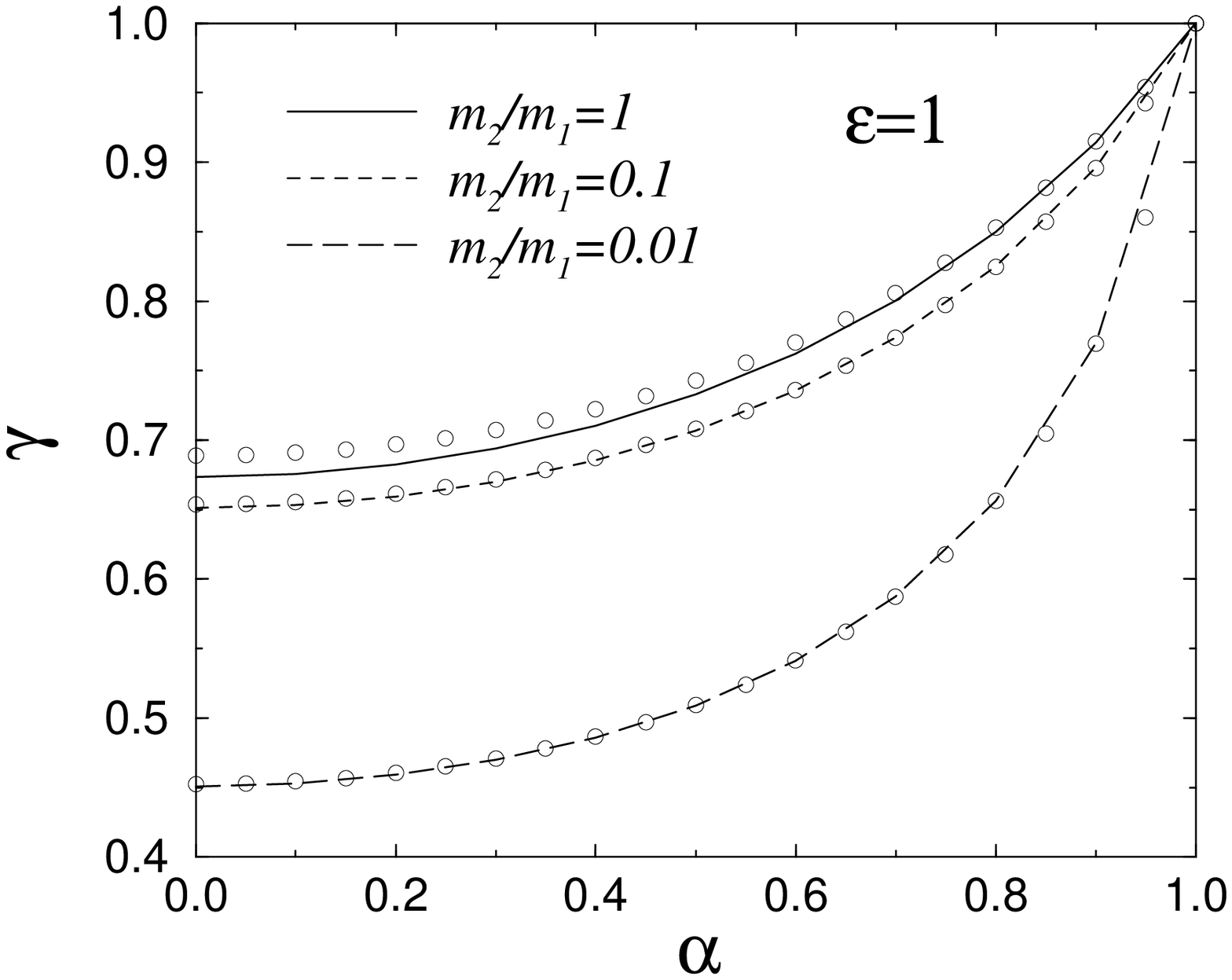,width=6cm,angle=0}
\psfig{figure=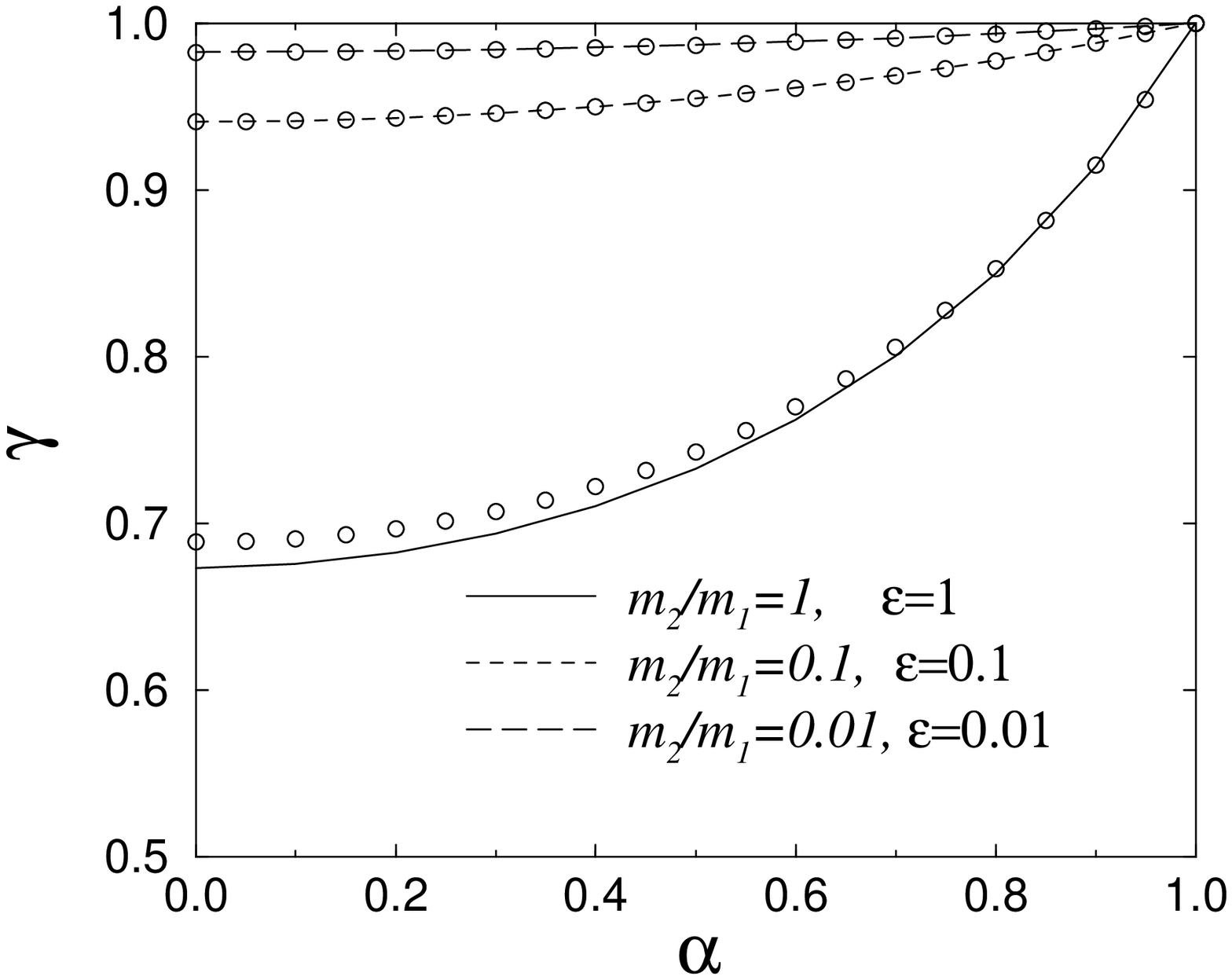,width=6cm,angle=0}}
\caption{Comparison of the simulation results found in \cite{biben} (lines) with the
solution of equation (\ref{eq:biben}) (circles), for various values of the mass ratio
and of the parameter $\epsilon$ recalled in (\ref{eq:biben}).}
\label{fig:biben}
\end{figure}

In Fig. \ref{fig:biben}, 
the solution of equation (\ref{eq:biben}) is compared to the results
reported in \cite{biben}.
The agreement is excellent, which may be traced back to the
analysis of \cite{biben}, showing that the distribution function
$\Phi_1$ is very close to a Gaussian, although mathematically different 
(on the other hand and by definition of the model, $\Phi_2$ is strictly 
speaking a Gaussian).
The slight discrepancy obtained
at low $\alpha$ for $m_1=m_2$ corresponds to values of the
parameters for which the deviation of $\Phi_1$ from a Gaussian is stronger.

\paragraph{}
Finally, a forcing term
${\cal F}f_i({\bm v}) =\zeta \frac{\partial}{\partial {\bm v}} \cdot
[ {\bm v} f({\bm v}) ] $, which provides an Enskog-Boltzmann equation
formally equivalent to the free cooling case (see e.g. \cite{Montanero}),
leads back to the results of \cite{garzo} obtained in this situation: 
the term ${\cal FT}_i$ is indeed proportional to
$T_i$, so that writing $\partial_t T_i =0$ in (\ref{eq:cooling})
yields the same equation for $\gamma$ as
equating the two cooling rates 
$\partial_t T_i / T_i$ when ${\cal FT}_i = 0$.

\section{The stochastic thermostat}
\label{sec:3}

In this section, we consider the situation of energy supply through
random kicks
\cite{Williams,Puglisi,twan,Pre1,Montanero,Cafiero,Moon,Pre2,Garzo}: 
the particles are submitted between collisions to an 
uncorrelated white noise (e.g. Gaussian). The equation of motion for a particle is
then
\begin{equation}
m_i \frac{d {\bm v}}{dt} = {\bm F}_i + m_i { \widehat{\bm\xi}}_i
\end{equation}
where ${\bm F}_i$ is the force due to inelastic collisions, and
$\langle \xi_{i\alpha}(t) \xi_{j\beta}(t') \rangle =
\xi_0^2 \delta_{ij}\delta_{\alpha\beta} \delta(t-t')$, where Greek 
indices refer to Cartesian coordinates.
The associated forcing term in the Enskog equation is
$$
{\cal F}f=
\frac{\xi_0^2}{2} \left(\frac{\partial}{\partial {\bm v}} \right)^2
f({\bm v},t) \ ,
$$
so that ${\cal FT}_i = m_i \xi_0^2$. We do not claim that the forcing term
considered here is the most suited to describe vibro-fluidized beds,
but it mimics an important effect of energy injection by a moving
piston: in experiments, particles undergoing collisions with the piston
(of large mass) 
gain a velocity that is decorrelated from their masses, so that more
kinetic energy is injected into the population of large mass.

The corresponding 
equation for $\gamma=T_1/T_2$ reads:
\begin{eqnarray}
& &\chi_{11} \,\sigma_{11}^{d-1} (1-\alpha_{11}^2) \frac{n_1}{n_2}
\left(\frac{m_2}{m_1}\right)^{3/2} \gamma^{3/2} 
+\sqrt{2} \, \chi_{12} \,\sigma_{12}^{d-1}
\left[ (1-\alpha_{12}^2)
\left( \mu_{21}^2 - \frac{n_1}{n_2} \mu_{12}^2 \right)
\left( 1+ \frac{m_2}{m_1} \gamma \right)^{3/2} \right.\nonumber\\
&+&2 \mu_{21} (1 + \alpha_{12} )
\left. \left(\mu_{21} +  \frac{n_1}{n_2} \mu_{12} \right)
\left( 1+ \frac{m_2}{m_1} \gamma \right)^{1/2}(\gamma-1) \right]
\,=\, \chi_{22}\, \sigma_{22}^{d-1} (1-\alpha_{22}^2) \ .
\label{eq:gamma}
\end{eqnarray}
The temperature ratio $\gamma$ therefore depends in a non-trivial 
way on the ratios of masses, densities and diameters, and also
on the inelasticities $\alpha_{ij}$ and pair correlation functions 
$\chi_{ij}$. It may be checked that in the limit of vanishing
inelasticities, $\gamma \to 1$ as it should. Moreover, for 
mechanically equivalent particles 
(i.e. $m_1=m_2$, $\sigma_1=\sigma_2$ and $\alpha_{11}=\alpha_{22}=\alpha_{12}$),
we should also recover equipartition ($\gamma=1$) irrespective of densities.
This is the case in the Boltzmann limit (low densities where all pair
correlation functions $\chi_{ij} \to 1$). At arbitrary packing fraction,
the various approximation for the $\chi_{ij}$ that may be found in the literature
\cite{Lee,Santos} are such that $\chi_{ij}$ no longer depends on $i$ and $j$
when $\sigma_1=\sigma_2$, so that equipartition holds for mechanically 
equivalent particles.

Since equation (\ref{eq:gamma}) relies on a Gaussian approximation for
$\Phi_i$, we have compared our approach to the results of Monte Carlo
simulations (the so-called DSMC technique \cite{Bird}) where the
non-linear Boltzmann equation is solved numerically for both
species. As we solve numerically the {\em homogeneous} Boltzmann
equation, the phenomena of segregation or clustering are explicitly
discarded.

In the following sections, we will study more precisely some cases
that could have experimental relevance, and for the sake of
simplicity, we considered $\chi_{ij}=1$. All the results are given for
the three dimensional case; note however that for $\sigma_1=\sigma_2$
the temperature ratio becomes $d$-independent.

\subsection{Equal inelasticities: $\alpha_{ij}=\alpha$}

\begin{figure}[htb]
\centerline{
\psfig{figure=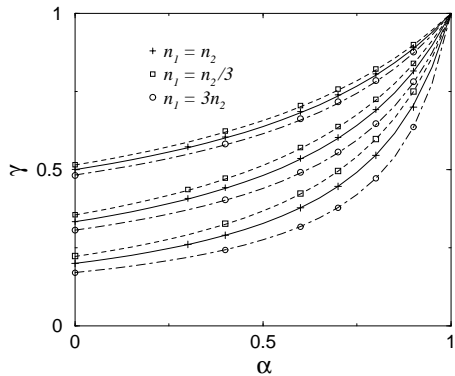,width=6cm,angle=0}}
\caption{Temperature ratio $\gamma=T_1/T_2$ 
as a function of inelasticity, for
$\alpha_{11} = \alpha_{22} = \alpha_{12}=\alpha$, and grains
of equal radii $(\sigma_1=\sigma_2$). The curves show
the solutions of equation (\ref{eq:gamma}) whereas the symbols display
the results of DSMC simulations. The top three curves correspond to 
a mass ratio $m_2/m_1=2$, the three intermediate ones to
$m_2/m_1=3$ and $m_2/m_1=5$ for the three bottom curves. For each mass ratio,
several density ratios have been considered:
$n_2/n_1=3$ (squares and dashed lines), $n_2/n_1=1$
(pluses and continuous lines), and $n_2/n_1=1/3$ (circles and
dot-dashed lines).
}
\label{fig:samealpha}
\end{figure}

We first consider the case of equal restitution coefficients
($\alpha_{ij}=\alpha$), for materials having similar elastic
properties.  The dependence of $\gamma$ on the mass and number density
ratios, for equal sizes, is shown in Figure \ref{fig:samealpha}.
Excellent agreement is found between DSMC simulations (symbols) and
the solution of equation (\ref{eq:gamma}). It turns out indeed that
the velocity distributions measured in Monte Carlo simulations are
very close to Gaussians. As may be expected, $\gamma$ is a decreasing
function of $m_2/m_1$. The density dependence is relatively weak (the
temperature ratio slightly increases when $n_2/n_1$ increases by an
order of magnitude).

\begin{figure}[htb]
\centerline{
\psfig{figure=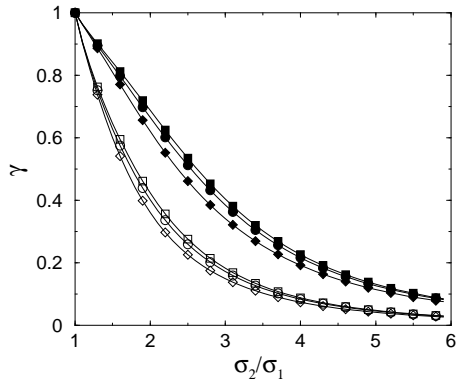,width=6cm,angle=0}}
\caption{Temperature ratio from eq (\ref{eq:gamma})
for grains made of the same material, i.e.
for $m_2/m_1=(\sigma_2/\sigma_1)^3$, and $\alpha_{ij}=\alpha$.
Filled symbols correspond to $\alpha=0.9$, open ones to
$\alpha=0.7$. The squares are for $n_2/n_1=3$, the circles for $n_2/n_1=1$ 
and the diamonds for $n_2/n_1=1/3$.
}
\label{fig:samematerial}
\end{figure}

We have also considered two types of beads of the same material, i.e.
the same restitution coefficient $\alpha_{11}=\alpha_{22}=\alpha_{12}$
and same mass density $\rho$: the ratio of masses $m_2/m_1$ is then
equal to $(\sigma_2/\sigma_1)^3$ for three-dimensional beads. Figure
\ref{fig:samematerial} shows a strong influence of the size ratio, for
two experimentally relevant values of $\alpha$: $\gamma$ decreases
very strongly as soon as $\sigma_2$ is two or three times $\sigma_1$.
Once again, the number density ratio has a relatively small incidence
on $\gamma$. It is interesting to disentangle the effects of
$\sigma_2/\sigma_1$ and $m_2/m_1$, by varying one parameter alone, the
other being kept constant. It appears that the leading effect in the
decrease of $\gamma$ observed in Figure \ref{fig:samematerial} is
ascribable to a change in mass ratio, and not in size: the results
obtained at $\sigma_1=\sigma_2$ varying $m_2/m_1$ are close to those
reported in Fig. \ref{fig:samematerial}, but surprisingly give a lower
$\gamma$ (e.g. the results displayed in Figure \ref{fig:samematerial}
for $\alpha=0.9$ and equal densities are $\gamma = 0.66$ and $0.38$
for $\sigma_2/\sigma_1=2$ and $3$ respectively, whereas with
$\sigma_1=\sigma_2$, we obtain $\gamma= 0.59$ for $m_2/m_1=2^3$ and
$\gamma=0.29$ for $m_2/m_1=3^3$).

\subsection{Comparison with experiments}

For glass spheres with size ratio $\sigma_2/\sigma_1 = 1.25$, Wildman
and Parker have measured a temperature ratio $\gamma = T_1/T_2$ in the
range 0.75-0.8 \cite{Wildman}, with a weak dependence on densities
(except may be in the limit of large grains predominance where $n_2
\gg n_1$). Estimating the relevant restitution coefficient to be
$\alpha \simeq 0.9$ \cite{Wildman}, we obtain from
Eq. (\ref{eq:gamma}) $\gamma \simeq 0.9$ (see also
Fig. \ref{fig:samematerial}), with also a weak dependence on
$n_2/n_1$. It is however noteworthy that this weak dependence is
opposite to that observed experimentally: when the proportion of large
grains is increased, we obtain an increase of $\gamma$. For
comparison, the temperature ratio obtained for the same parameters in
the homogeneous cooling stage \cite{garzo} is $\gamma \simeq 0.96$ and
the authors of \cite{Wildman} proposed a simplified theory for which
$\gamma$ is in the range 0.4-0.7.

\begin{figure}[htb]
\centerline{
\psfig{figure=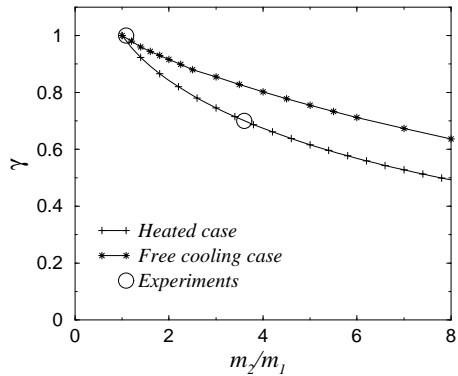,width=6cm,angle=0}}
\caption{$T_1/T_2$ as a function of mass ratio, for 
$\alpha_{ij}=0.85$ and $\sigma_1=\sigma_2$, for the stochastic thermostat and 
for the free cooling regime, together with the experimental
data of \cite{menon}. Note that the values $\alpha_{ij}=0.85$ are only 
schematic and cannot be intended as an exact description of
the experimental situation.
}
\label{fig:a.85}
\end{figure}

The results obtained by Feitosa and Menon \cite{menon} confirm the
very weak influence of $n_2/n_1$ when the grains (of equal size) are
made of two different materials. For a mixture glass/aluminum with
mass ratio $m_2/m_1 = 1.09$, $\gamma$ is measured very close to 1,
whereas for a more asymmetric mixture of glass and brass with $m_2/m_1
\simeq 3.6$, $\gamma \simeq 0.7$. Making use of equation
(\ref{eq:gamma}) with schematic inelasticities
$\alpha_{11}=\alpha_{12}=\alpha_{22}=0.85$, we obtain $\gamma= 0.98$
for $m_2/m_1=1.09$ and $\gamma = 0.7$ for $m_2/m_1=3.6$.  In the free
cooling regime, the corresponding ratios are 0.99 and 0.82. These
results are displayed in Fig.~\ref{fig:a.85}.

\begin{figure}[htb]
\centerline{
\psfig{figure=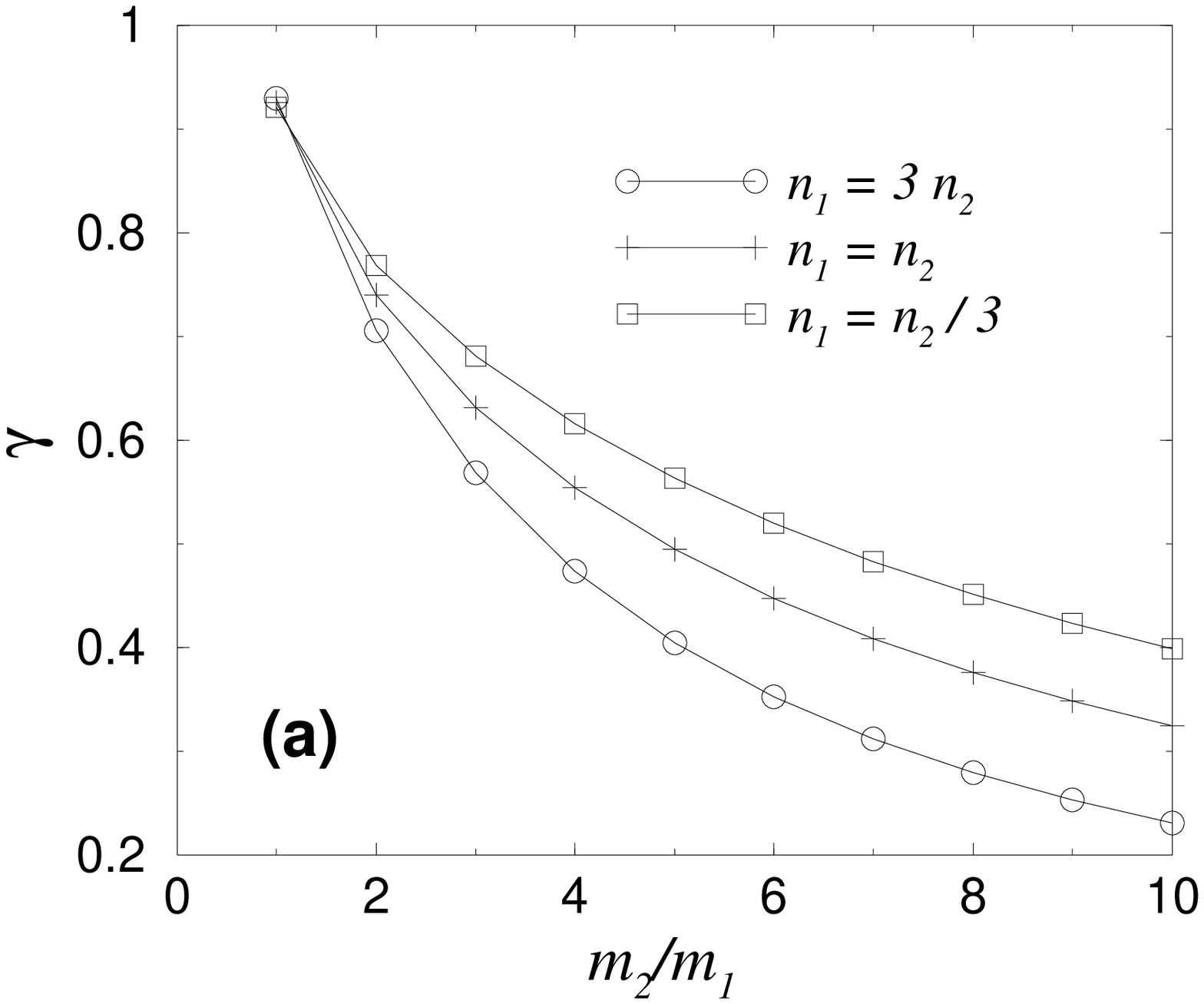,width=6cm,angle=0}
\psfig{figure=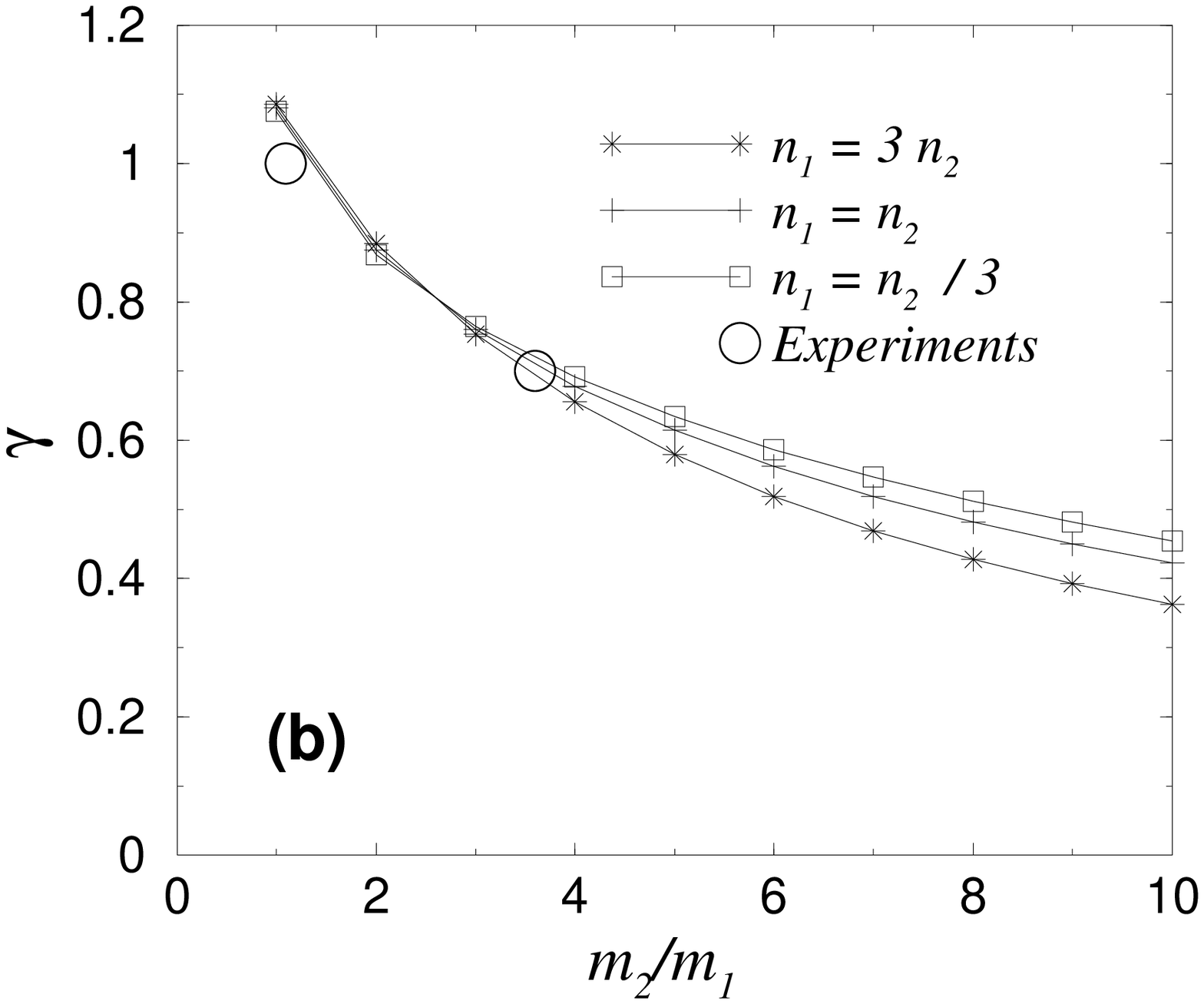,width=6cm,angle=0}}
\caption{(a): $T_1/T_2$ as a function of mass ratio, for 
$\alpha_{11}=0.7$, $\alpha_{12}=0.8$, 
$\alpha_{22}=0.9$ and $\sigma_1=\sigma_2$.\\
(b): same with ``reversed'' inelasticities 
($\alpha_{11}=0.9$, $\alpha_{12}=0.8$ and $\alpha_{22}=0.7$), together
with the experimental values of \cite{menon}.
}
\label{fig:a.7.8.9}
\end{figure}

\begin{figure}[htb]
\centerline{
\psfig{figure=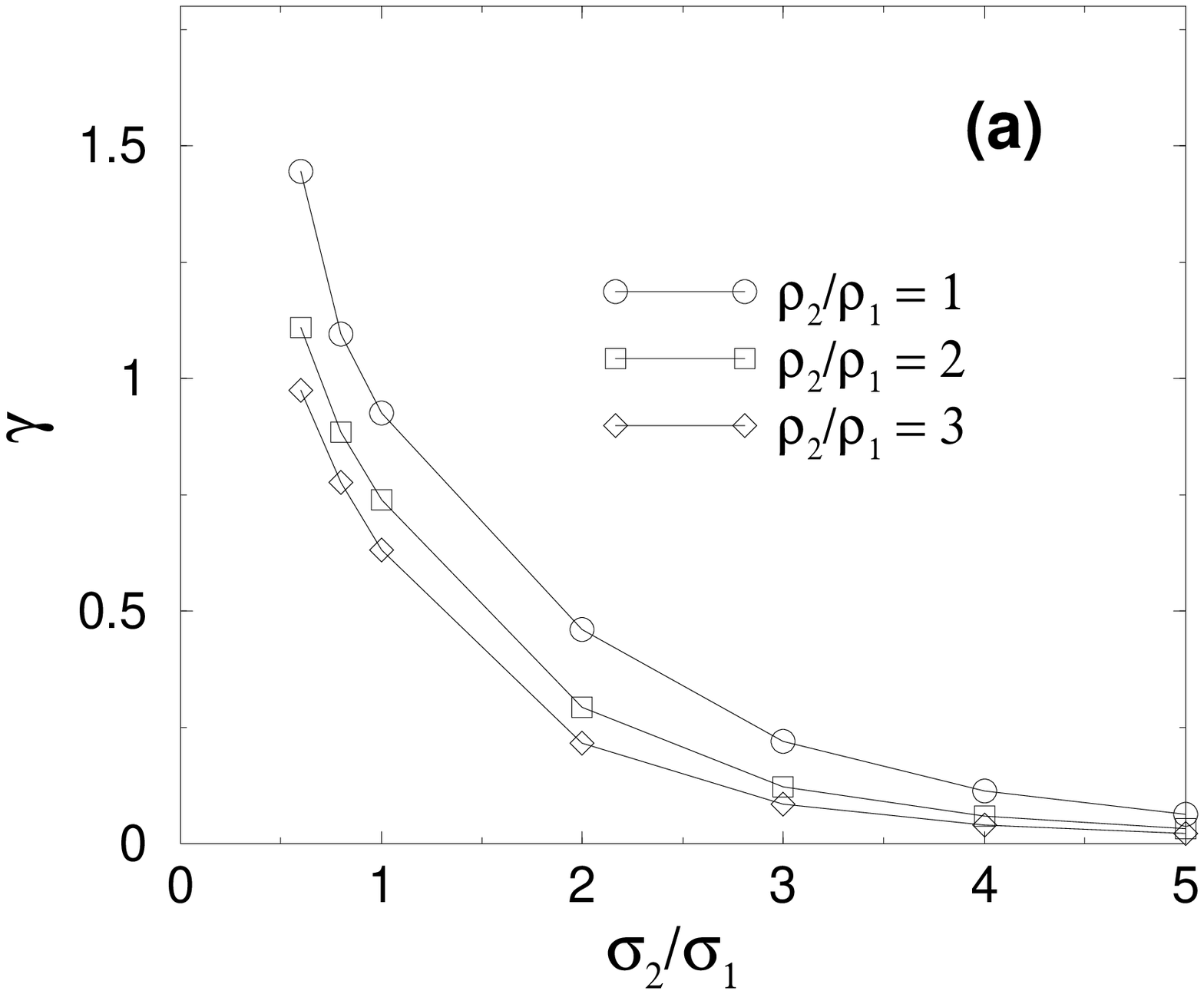,width=6cm,angle=0}
\psfig{figure=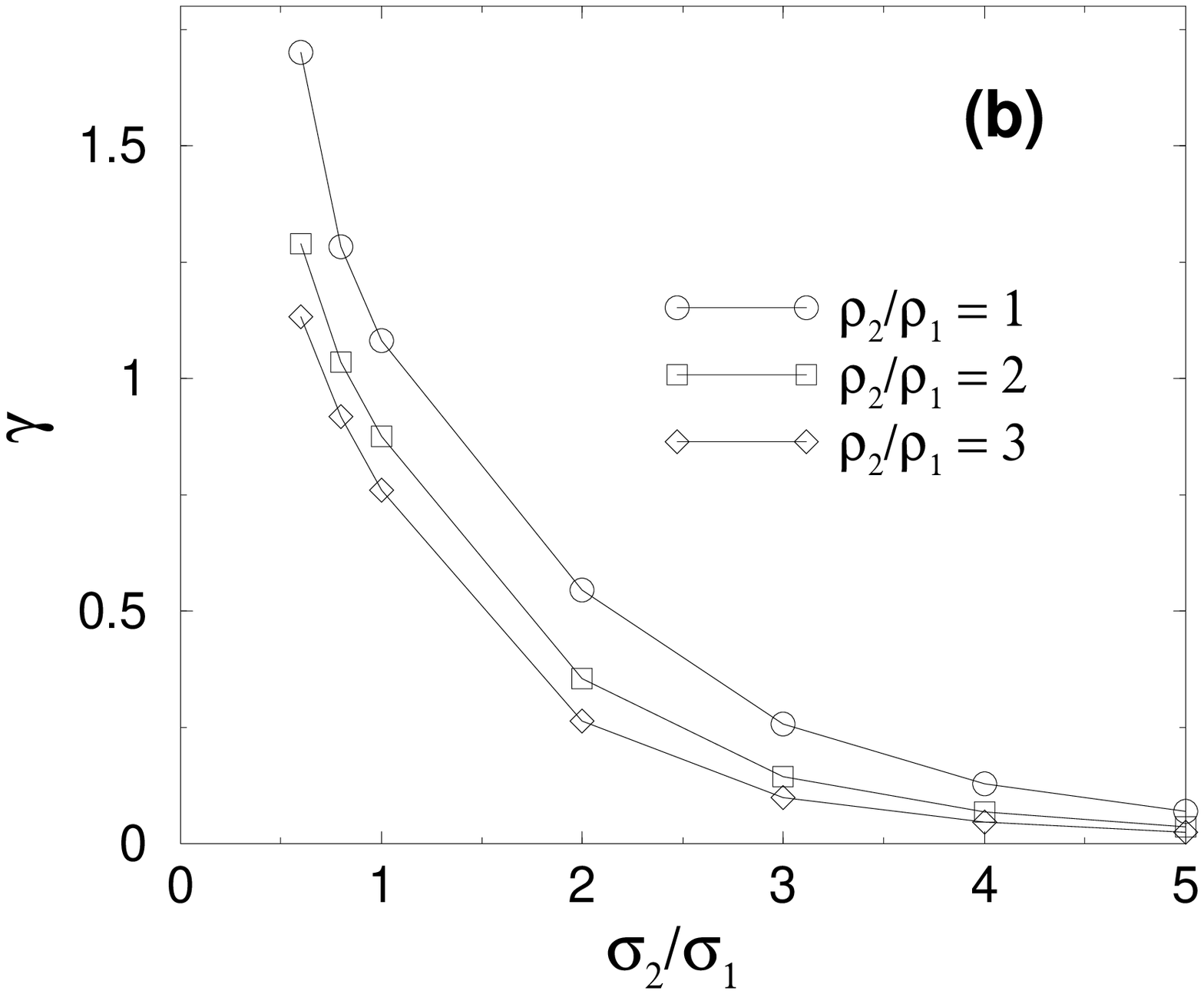,width=6cm,angle=0}}
\caption{(a): $\gamma$ versus size ratio for an equimolar mixture
($n_2=n_1$) with
$\alpha_{11}=0.7$, $\alpha_{12}=0.8$, 
$\alpha_{22}=0.9$ and different mass density ratios $\rho_2/\rho_1$.
 \\
(b): same with $\alpha_{11}=0.9$, $\alpha_{12}=0.8$, 
$\alpha_{22}=0.7$ and again $n_2=n_1$.
}
\label{fig:a.7.8.9rm}
\end{figure}

Other results are given in Figures 
\ref{fig:a.7.8.9} and \ref{fig:a.7.8.9rm}, where the partial inelasticities
are not taken equal, but are given values that we expect to be of
experimental relevance: the experimental data of
\cite{menon} are therefore also reported in Fig. \ref{fig:a.7.8.9}(b).
In Fig. \ref{fig:a.7.8.9}, the sizes
of the particles are taken equal and the mass ratio changes,
while Fig. \ref{fig:a.7.8.9rm} displays the influence of the size ratio
when the density ratios are fixed.

The situation reported in \cite{menon}
corresponds to that of the Figures \ref{fig:a.7.8.9}b
and \ref{fig:a.7.8.9rm}b, where the heavier grains are also the more
dissipative.  As may be observed in 
Fig. \ref{fig:a.7.8.9}b, a variation of $n_1/n_2$ from 1/3 to 3 leaves
$\gamma$ roughly unaffected for $m_2/m_1 \leq 3$.

It may be noted that $\gamma$ is not bounded from above by 1, and
values slightly above 1 are obtained even when $m_2>m_1$ by
conveniently choosing the inelasticities (or, at fixed inelasticities
and densities, by conveniently choosing the sizes).  $\gamma$ in
nevertheless generically smaller than $1$ for $m_2 > m_1$: the heavier
particles have a larger kinetic energy.

\section{Conclusion}

We have considered heated binary granular mixtures from the
point of view of kinetic theory. As in the free cooling case,
and in agreement with recent experimental data, the granular
temperatures of the components of the mixture differ. 
This finding is not surprising in a non-equilibrium system,
where the``temperature'' does not have any thermodynamical 
relevance.

Using a mean-field approach with the assumption of isotropic Gaussian velocity
distributions, we
have derived an equation for the temperature ratio $\gamma$ that
may be adapted for various kinds of heating mechanisms, and easily solved
once the controlling parameters have been chosen.
In particular, the values obtained within the stochastic thermostat framework
are compatible with those measured in the experiments reported in 
\cite{Wildman,menon}. Even if a quantitative comparison with experiments is somehow
pointless given the simplicity of our approach,
similar trends are observed. For example, the heavier particles
carry generically more kinetic energy than the lighter ones,
the ratio being insensitive to the relative number fraction of both species.
It also appears that the breakdown of energy equipartition is all the
more pronounced as the mass ratio is increased, the size ratio
playing only a minor role.

Acknowledgments: We would like to thank T. Biben, R.D. Wildman and 
C.M. Hrenya for communicating their results prior to publication.

\appendix

\section*{Appendix}
In this appendix we show how to perform the integrals over 
$\widehat{\bm \sigma}$, in order to obtain equation (\ref{eq:cooling}).
We start from the identity
\begin{equation}
\int  d\bm{v} v^2 J_{ij}[ \bm{v} | f_i, f_j] \,=\,
\chi_{ij} \sigma_{ij}^{d-1}
\int d\bm{v}_1 d\bm{v}_2 \int' d\widehat{\bm{\sigma}} 
(\widehat{\bm{\sigma}}\cdot \bm{v}_{12}) 
 f_i(\bm{v}_1)f_j(\bm{v}_2) (v_1''^2 - v_1^2) \ ,
\end{equation}
with $\bm{v}_1'' = \bm{v_1} - \mu_{ji} (1+\alpha_{ij})
(\widehat{\bm{\sigma}}\cdot \bm{v}_{12}) \widehat{\bm{\sigma}}$,
i.e. where
$$
v_1''^2 - v_1^2 = \mu_{ji}^2 (1+\alpha_{ij})^2
(\widehat{\bm{\sigma}}\cdot \bm{v}_{12})^2 
-2\mu_{ji}(1+\alpha_{ij})(\widehat{\bm{\sigma}}\cdot \bm{v}_{12})
(\widehat{\bm{\sigma}}\cdot \bm{v}_1) \ .
$$
Using the unit vector $\widehat{\bm{c}}_{12}=\bm{v}_{12}/v_{12}$, and
the known integrals
$\beta_n = \int' d\widehat{\bm{\sigma}} 
(\widehat{\bm{\sigma}}\cdot \bm{c}_{12})^n$ (see e.g. \cite{twan}),
the first term is readily computed and yields:
\begin{equation}
\beta_3
\chi_{ij} \sigma_{ij}^{d-1}  \mu_{ji}^2 (1+\alpha_{ij})^2
\int d\bm{v}_1 d\bm{v}_2  f_i(\bm{v}_1)f_j(\bm{v}_2) 
v_{12}^3 \ .
\end{equation}
To compute the term containing $(\widehat{\bm{\sigma}}\cdot \bm{v}_1)$,
we choose one of the unit vectors to be along
$\bm{v}_{12}$, and decompose:
\begin{equation}
\bm{v}_{12}= v_{12} \,\widehat{\bm{e}}_1, \qquad
\bm{v}_1 = \frac{\bm{v}_1 \cdot \bm{v}_{12}}{v_{12}} \,\widehat{\bm{e}}_1
+ \bm{v}_1^\perp ,  \qquad
\widehat{\bm{\sigma}} = \frac{\widehat{\bm{\sigma}} \cdot \bm{v}_{12}}{v_{12}} \,
\widehat{\bm{e}}_1 + \widehat{\bm{\sigma}}^\perp \ .
\end{equation}
$(\widehat{\bm{\sigma}}\cdot \bm{v}_1)$ is then written as
$$
\frac{(\bm{v}_1 \cdot \bm{v}_{12})
(\widehat{\bm{\sigma}} \cdot \bm{v}_{12})}{v_{12}^2} + 
\widehat{\bm{\sigma}}^\perp \cdot  \bm{v}_1^\perp \ ,
$$
and the term $\widehat{\bm{\sigma}}^\perp \cdot  \bm{v}_1^\perp$ gives
a vanishing contribution in the integral over $\widehat{\bm{\sigma}}$ for
symmetry reasons. We are therefore left with
$$
\int' d\widehat{\bm{\sigma}} 
\frac{(\bm{v}_1 \cdot \bm{v}_{12})
(\widehat{\bm{\sigma}} \cdot \bm{v}_{12})^3}{v_{12}^2} = v_{12} \beta_3
(\bm{v}_1 \cdot \bm{v}_{12}) \ .
$$
Rearranging terms and writing $\bm{v}_1=\mu_{ji}\bm{v}_{12}
+\mu_{ij} \bm{v}_1 + \mu_{ji}\bm{v}_2=\mu_{ji}\bm{v}_{12} +
\bm{V}_{ij}$, one finally obtains 
 equation (\ref{eq:cooling}).



\begin{thebibliography}{99}

\bibitem{Jaeger} H.M. Jaeger, S.R. Nagel and R.P. Behringer, Rev. Mod.
Phys. {\bf 68}, 1259 (1996); L.P. Kadanoff, Rev. Mod. Phys. 
{\bf 71}, 435 (1999). 

\bibitem{Duparcmeur}
Y. Limon Duparcmeur, Th\`ese de l'universit\'e de Rennes I (1996).

\bibitem{garzo}
V. Garz\'o and J. Dufty,
Phys. Rev. E {\bf 60} 5706 (1999).

\bibitem{Huilin}
L. Huilin, L. Wenti, B. Rushan, Y. Lidan and D. Gidaspow,
Physica A {\bf 284}, 265 (1999). 

\bibitem{losert}
W. Losert, D.G.W. Cooper, J. Delour, A. Kudrolli and J.P. Gollub,
Chaos {\bf 9}, 682 (1999).

\bibitem{Wildman}
R.D. Wildman and D.J. Parker, Phys. Rev. Lett. {\bf 88}, 064301 (2002).

\bibitem{menon} 
K. Feitosa and N. Menon, preprint, cond-mat/0111391.

\bibitem{MontaneroHCS}
J. M. Montanero and V. Garz\'o, Gran. Matter {\bf 4}, 17 (2002). 

\bibitem{Clelland}
R. Clelland and C. M. Hrenya, Phys. Rev. E {\bf 65}, 031301 (2002). 

\bibitem{MontaneroShear}
J. M. Montanero and V. Garz\'o, preprint cond-mat/0201175.

\bibitem{biben}
T. Biben, Ph. A. Martin and J. Piasecki, preprint.

\bibitem{puglisi2}
U. Marini Bettolo Marconi and A. Puglisi, cond-mat/0112336 Phys. Rev. E (2002)
and cond-mat/0202267.

\bibitem{martin}
Ph. A. Martin and J. Piasecki, Europhys. Lett. {\bf 46}, 613 (1999).

\bibitem{Williams}
D.R.M. Williams and F.C. MacKintosh, Phys. Rev E {\bf 54}, R9 (1996).
 
\bibitem{Puglisi} 
A. Puglisi, V. Loreto, U. Marini Bettolo Marconi and A. Vulpiani, 
Phys. Rev. E {\bf 59}, 5582 (1999).
 
\bibitem{twan}
T.P.C. van Noije and M.H. Ernst,  Gran. Matter {\bf 1}, 57 (1998).
 
\bibitem{Pre1}
T.P.C. van Noije, M.H. Ernst, E. Trizac and I. Pagonabarraga,
Phys. Rev. E {\bf 59}, 4326 (1999).
     
\bibitem{Montanero}
J.M. Montanero and A. Santos, Granular Matter {\bf 2}, 53 (2000). 

\bibitem{Cafiero}
R. Cafiero, S. Luding and H.J. Herrmann, 
Phys. Rev. Lett. {\bf 84}, 6014 (2000).

\bibitem{Moon}
S.J. Moon, M.D. Shattuck and J.B. Swift, Phys. Rev. E {\bf 64}, 031303 (2001). 

\bibitem{Pre2}
I. Pagonabarraga, E. Trizac, T.P.C. van Noije, and M.H. Ernst,
Phys. Rev. E {\bf 65}, 011303 (2002).

\bibitem{Garzo}
V. Garz\'o and J.M. Montanero, cond-mat/0112241.

\bibitem{Landau}
L. Landau and E. Lifshitz, {\it Physical Kinetics}, Pergamon Press (1981).

\bibitem{Lee}
L.L. Lee and D. Levesque, Mol. Phys. {\bf 24}, 269 (1972). 

\bibitem{Santos}
A general procedure to infer the pair correlation functions in a multi-component
$d$-dimensional hard-sphere fluid from the equation of state of the monodisperse
system has been proposed by A. Santos, S.B. Yuste
and M. L\'opez de Haro, Mol. Phys. {\bf 96}, 1 (1999). 


\bibitem{Bird}
G. Bird, ``Molecular Gas Dynamics'' (Oxford University Press, New York, 1976)
and ``Molecular Gas Dynamics and the Direct Simulation of Gas flows''
(Clarendon Press, Oxford, 1994).


\end{thebibliography}
\end{document}